\documentclass[useAMS,usenatbib,usegraphicx]{mn2e}
\usepackage{times}

\title[Shrinkage estimation of the covariance matrix]
{Shrinkage estimation of the power spectrum covariance matrix}

\author[Pope \& Szapudi]{
Adrian C. Pope$^1$\thanks{E-mail: pope@ifa.hawaii.edu}
and Istv\'{a}n Szapudi$^1$\\
$^1$Institute for Astronomy, 2680 Woodlawn Drive, Honolulu, HI 96822}

\newcommand{\figtoy}{
  \begin{figure*}
    \begin{center}
      \includegraphics[width=\textwidth]{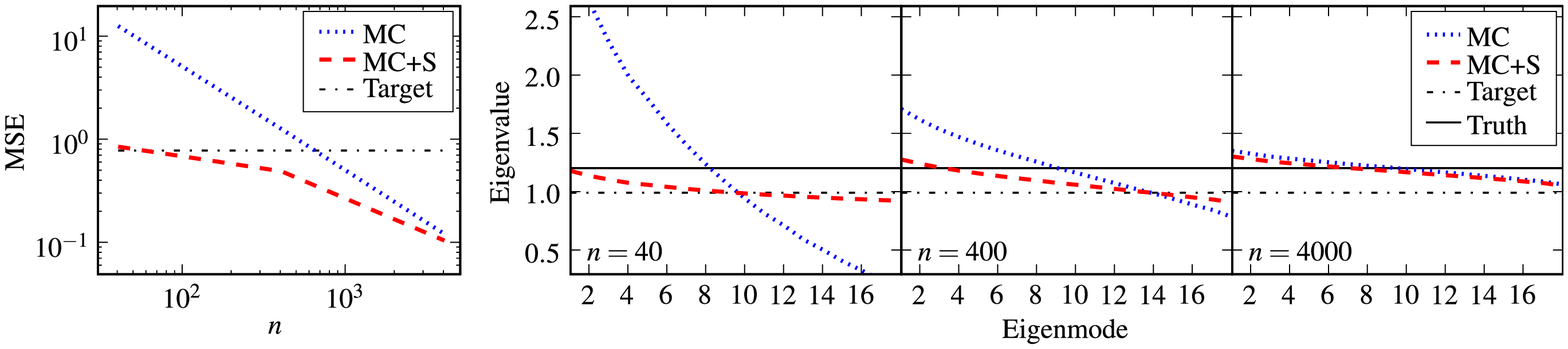}
    \end{center}
    \caption{
      Comparison of the Monte Carlo (MC) and shrinkage (MC+S) estimates 
      for the toy model covariance from Section~\ref{sub:toycov}.
      The plot at left shows the mean squared error (MSE) between the
      estimate and the known input covariance as a function of 
      the number of realizations, $n$.
      Results using only the target are also shown.
      The plots at right show the eigenvalue spectra for the covariance 
      estimators using different numbers of realizations, $n$.
      The known true eigenvalue spectrum and the eigenvalues of the target
      are also shown.
      All plots represent the averages of 100 simulations for each
      number of realizations, $n$.
    }
    \label{fig:toy}
  \end{figure*}
}

\newcommand{\figpk}{
  \begin{figure}
    \begin{center}
      \includegraphics[width=\columnwidth]{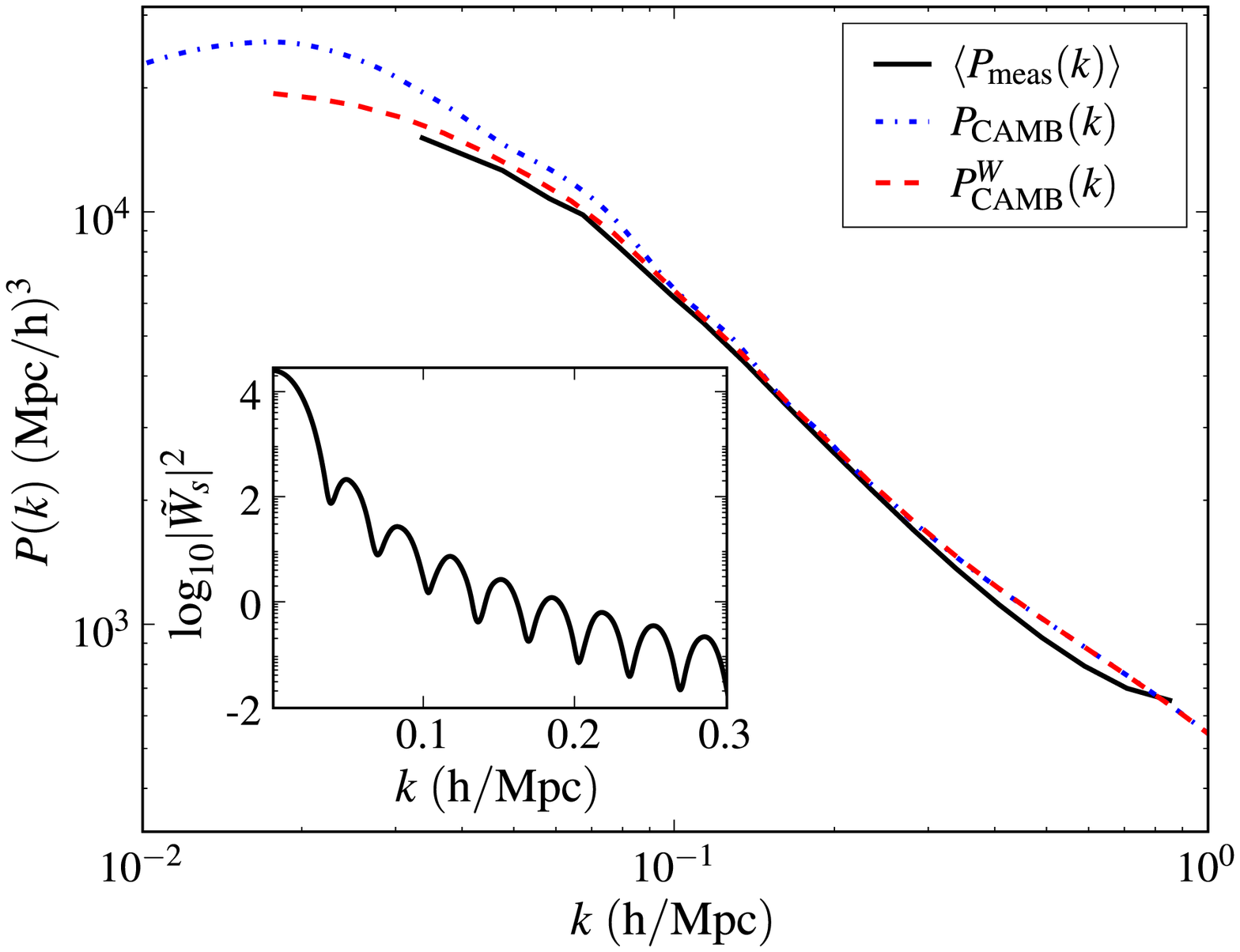}
    \end{center}
    \caption{Plot of the  input theoretical power spectrum 
      $\left( P_{\rm CAMB} \right)$, 
      the theoretical power spectrum convolved with the survey window function
      $\left( P^W_{\rm CAMB} \right)$,
      and the averaged power spectrum measured from all of the sub-volumes 
      $\left( \langle P_{\rm meas} \rangle \right)$.
      Inset shows the spherically averaged survey window function.}
    \label{fig:pk}
  \end{figure}
}

\newcommand{\figsmc}{
  \begin{figure}
    \begin{center}
      \includegraphics[width=\columnwidth]{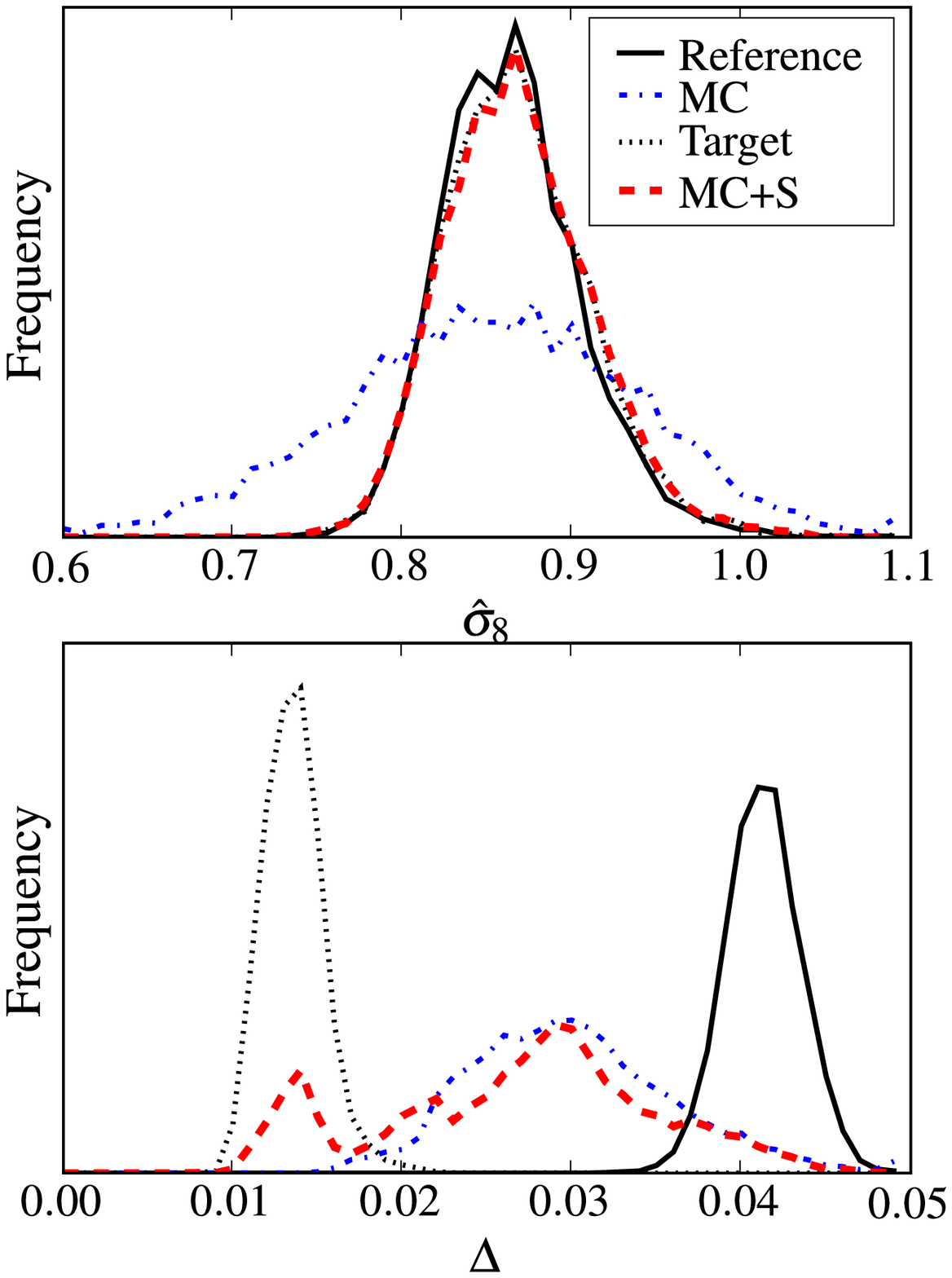}
    \end{center}
    \caption{(Area normalized) distributions of maximum-likelihood value, 
      $\hat{\sigma}_8$, and error bar, 
      $\Delta$, estimates for the Reference, Monte Carlo (MC), Monte
      Carlo Target only (Target), and the Monte Carlo + Shrinkage (MC+S)
      covariance matrix estimates.}
    \label{fig:s8mc}
  \end{figure}
}

\newcommand{\figcov}{
  \begin{figure}
    \begin{center}
      \includegraphics[width=\columnwidth]{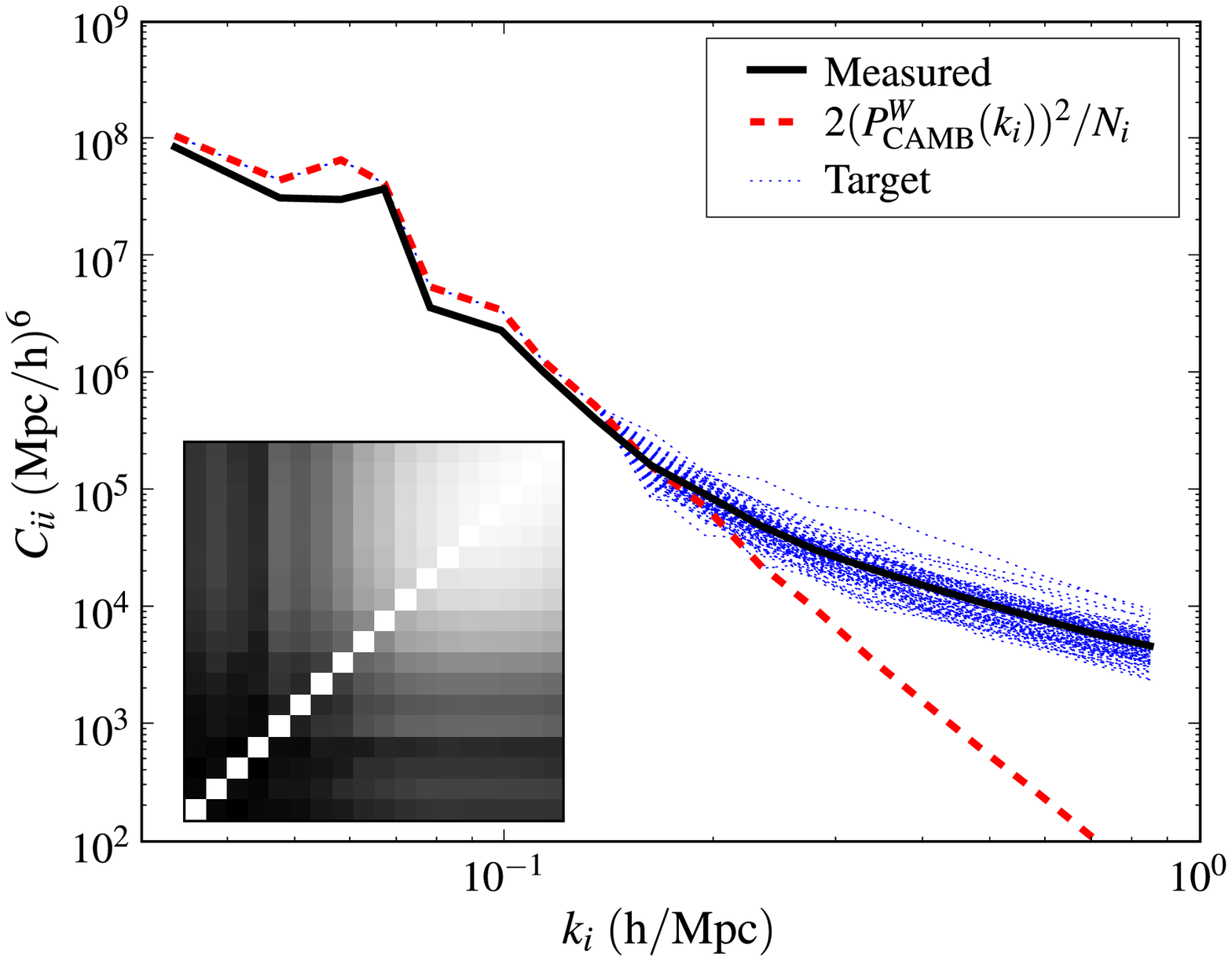}
    \end{center}
    \caption{Plot of the diagonal elements of the reference covariance
      matrix estimated from all of the sub-volumes, of a linear theory
      model for the covariance, and of the 102 target matrices for
      the Monte Carlo + Shrinkage estimates.
      Inset shows the reference correlation matrix in a linear stretch.}
    \label{fig:cov}
  \end{figure}
}

\newcommand{\figsjk}{
  \begin{figure}
    \begin{center}
      \includegraphics[width=\columnwidth]{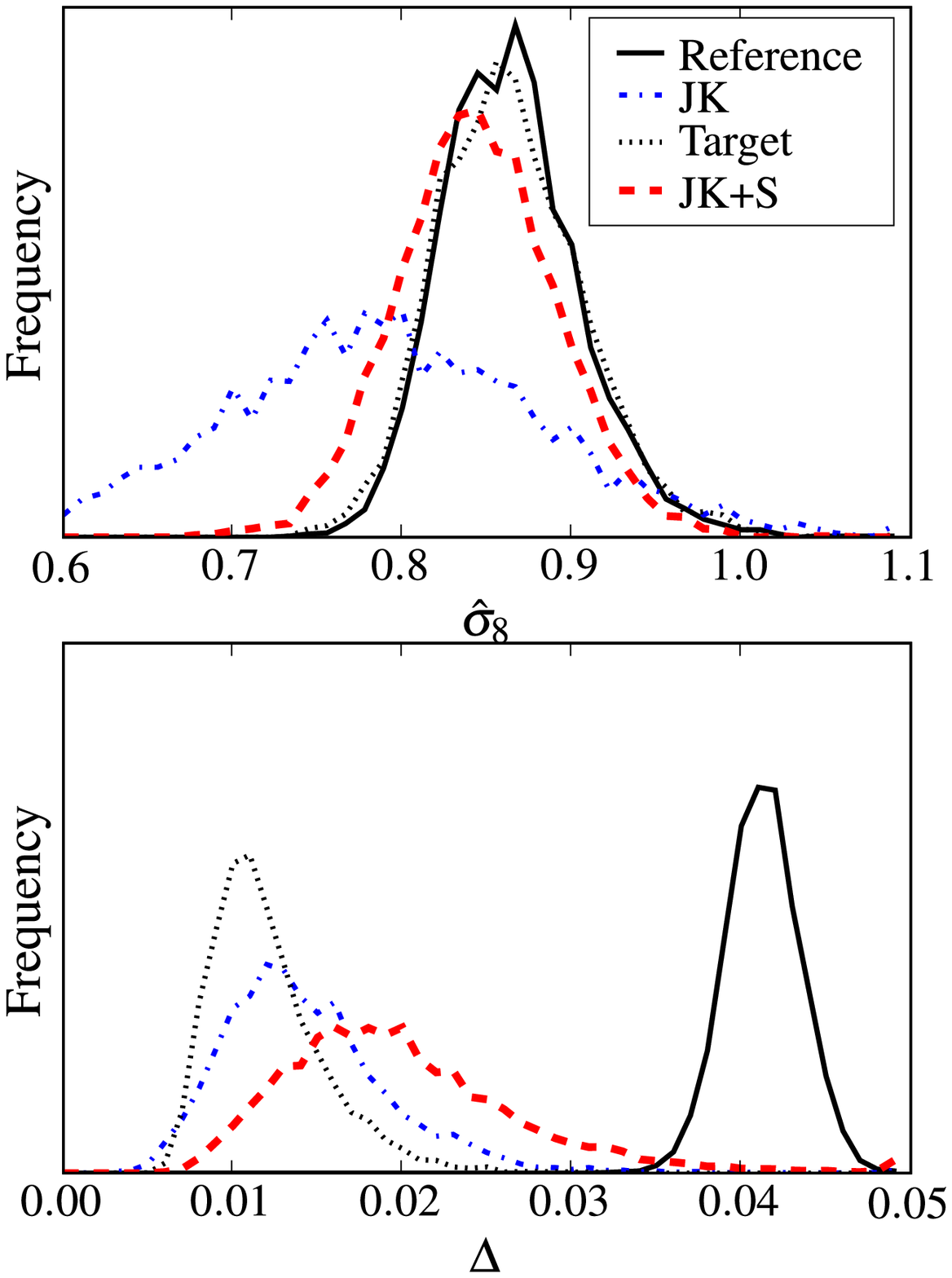}
    \end{center}
    \caption{(Area normalized) distributions of maximum-likelihood value, 
      $\hat{\sigma}_8$, and error bar, 
      $\Delta$, estimates for the Reference, Jackknife (JK), Jackknife
      Target only (Target), and the Jackknife + Shrinkage (JK+S)
      covariance matrix estimates.}
    \label{fig:s8jk}
  \end{figure}
}

\newcommand{\figeigen}{
  \begin{figure}
    \begin{center}
      \includegraphics[width=\columnwidth]{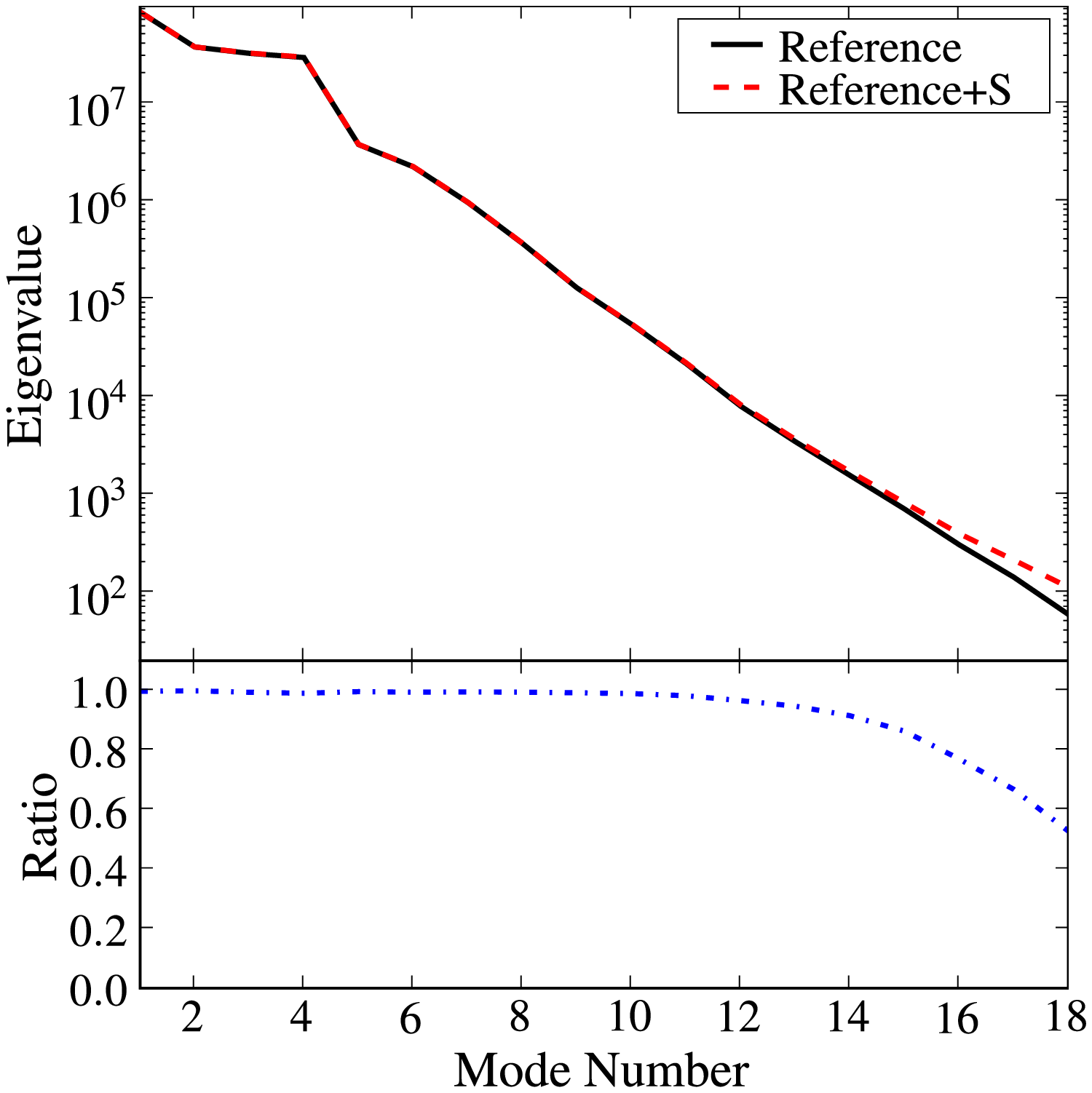}
    \end{center}
    \caption{Plot of the (sorted) eigenvalue spectrum for the Reference 
      covariance matrix, empirically estimated from 4096 sub-volumes, and the 
      shrinkage of that reference covariance matrix against our diagonal 
      target (Reference+S).  The lower panel shows the ratio of these 
      eigenvalue spectra.}
    \label{fig:eigen}
  \end{figure}
}

\newcommand{\figlambda}{
  \begin{figure}
    \begin{center}
      \includegraphics[width=\columnwidth]{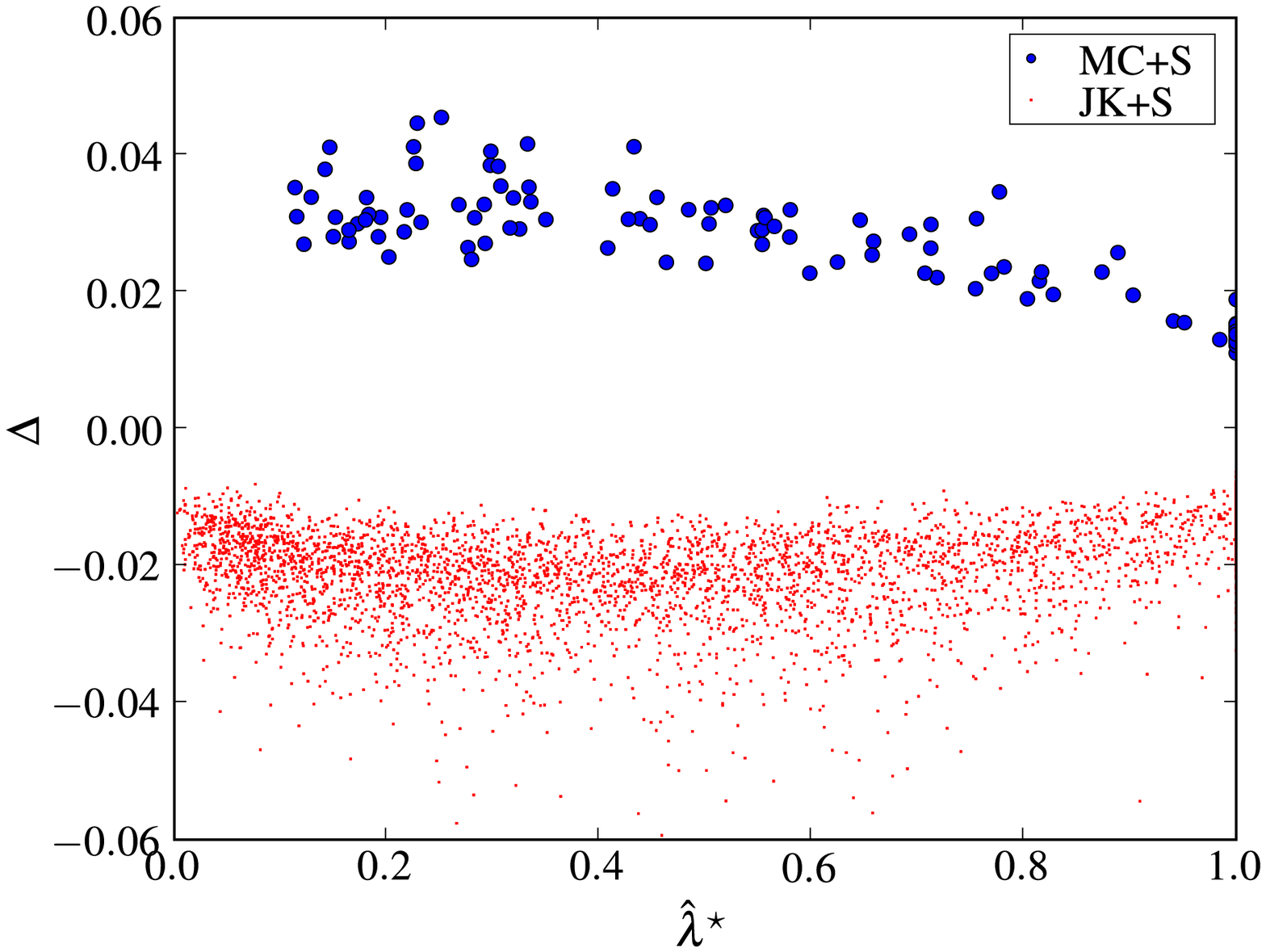}
    \end{center}
    \caption{Plot of estimated error bar, $\Delta$, as a function
      of the shrinkage intensity, $\hat{\lambda}^{\star}$, for the 
      shrinkage versions of the Monte Carlo (MC+S) and jackknife 
      (JK+S) methods.  For clarity all of the MC+S error bars are plotted
      as positive and all of the JK+S as negative.}
    \label{fig:lambda}
  \end{figure}
}

\newcommand{\tableresults}{
  \begin{table}
    \caption{Statistics of $\hat{\sigma}_8$ and error bar estimates.}
    \label{tab:s8}
    \begin{tabular}{lllll}
      \hline
      $\mathbfss{C}$ & 
      $\langle \hat{\sigma}_8 \rangle$ & 
      $\sigma_{\hat{\sigma}_8}$ & 
      $\langle \Delta \rangle$ & 
      $\sigma_{\Delta}$ \\
      \hline
      Reference & 0.870 & 0.041 & 0.042 & 0.002 \\
      Monte Carlo & 0.853 & 0.088 & 0.031 & 0.006 \\
      Monte Carlo Target Only & 0.870 & 0.042 & 0.014 & 0.002 \\
      Monte Carlo + Shrinkage & 0.872 & 0.042 & 0.027 & 0.008 \\
      Jackknife & 0.790 & 0.102 & 0.015 & 0.005 \\
      Jackknife Target Only & 0.869 & 0.044 & 0.013 & 0.003 \\
      Jackknife + Shrinkage & 0.850 & 0.047 & 0.021 & 0.007 \\	
      \hline
    \end{tabular}

    \medskip
    The mean and standard deviation of the estimates of the
    maximum-likelihood estimate, $\hat{\sigma}_8$, and the one-sigma
    error bar, $\Delta$, using different methods to estimate the
    covariance matrix.
    \end{table}     
}

\newcommand{\tabletoy}{
  \begin{table*}
    \begin{minipage}{\textwidth}
      \caption{Shrinkage estimation of the mean of a noisy vector.}
      \label{tab:toy}
      \begin{tabular}{lllll|lllllllll}
	\hline
	\multicolumn{5}{l}{Input}  & \multicolumn{9}{l}{Output} \\
	$\psi$ &
	$\sigma$ &
	$t$ &
	$n$ &
	$p$ &
	$\langle \lambda^{\star} \rangle$ &
	$\overline{\hat{\lambda}^{\star}}$ &
	$\Delta \hat{\lambda}^{\star}$ &
	$\langle {\rm MSE}(\bmath{u}^{\star}) \rangle$ &
	$\overline{{\rm MSE}(\bmath{u}^{\star})}$ &
	$\langle {\rm MSE}(\bmath{u}) \rangle$ &
	$\overline{{\rm MSE}(\bmath{u})}$ &
	$\langle {\rm MSE}(\bmath{t}) \rangle$ &
	$\overline{{\rm MSE}(\bmath{t})}$ \\
	\hline
	1.1 & 1.0 & 1.0 & 100 & 100 & 0.50 & 0.509 & 0.003 & 0.50 & 0.52 & 1.00 & 1.02 & 1.00 & 1.00\\
	1.2 & & & & & 0.20 & 0.205 & 0.0002 & 0.80 & 0.80 & 1.00 & 0.99 & 4.00 & 4.00\\
	& 0.9 & & & & 0.45 & 0.459 & 0.003 & 0.45 & 0.46 & 0.81 & 0.80 & 1.00 & 1.00\\
	& & 0.9 & & & 0.20 & 0.201 & 0.0003 & 0.80 & 0.80 & 1.00 & 1.00 & 4.00 & 4.00\\
	& & & 50 & & 0.67 & 0.696 & 0.008 & 0.67 & 0.71 & 1.00 & 2.00 & 1.00 & 1.00\\
	& & & & 50 & 0.50 & 0.529 & 0.01 & 0.25 & 0.26 & 0.50 & 0.49 & 0.50 & 0.50\\
	\hline
      \end{tabular}
      
      \medskip 
      Results of simulations to test shrinkage estimation of the mean of
      a noisy vector.  The first five columns are the input values for
      the simulations.  The first row gives the fiducial values and
      subsequent rows only indicate parameters that were varied.
      The remaining columns list analytically predicted 
      (indicated by $\langle \rangle$)
      and measured (where $\bar{a}$ and $\Delta a$ indicate the sample
      mean and standard deviation for $a$) 
      quantities from the outputs of the simulations.
      We used 100 simulations for each set of input parameters.
      See Section~\ref{sub:toymean} for an explanation of the parameters
      and quantities.

    \end{minipage}
  \end{table*}
}

\begin{document}
\maketitle

\begin{abstract}
We seek to improve estimates of the power spectrum covariance matrix
from a limited number of simulations by employing a novel statistical
technique known as shrinkage estimation.
The shrinkage technique optimally combines an empirical 
estimate of the covariance with a model 
(the {\it target}) to minimize the {\it total} mean squared error compared 
to the true underlying covariance.
We test this technique on N-body simulations and evaluate its performance
by estimating cosmological parameters.
Using a simple diagonal target, we show that the shrinkage estimator 
significantly outperforms both the empirical covariance and the target 
individually when using a small number of simulations.
We find that reducing noise in the covariance estimate is essential for 
properly estimating the values of cosmological parameters as well as their
confidence intervals.
We extend our method to the jackknife covariance estimator and again
find significant improvement, though simulations give better results.
Even for thousands of simulations we still find evidence that our method 
improves estimation of the covariance matrix.
Because our method is simple, requires negligible additional numerical 
effort, and produces superior results, we always advocate shrinkage 
estimation for the covariance of the power spectrum and other large-scale 
structure measurements when purely theoretical modeling of the
covariance is insufficient.
\end{abstract}


\section{Introduction}
\label{sec:intro}

Large-scale structure statistics, especially power spectra, provide precise 
constraints on cosmological theories.
Accurate measurements are now possible with large-volume surveys and 
advancing computational power.
However, the measured power spectrum is not the only required ingredient 
for estimating cosmological parameters; the covariance matrix also carries 
a great deal of information that is vital for properly estimating parameter 
values and their confidence intervals.
Observational effects such as the survey geometry, redshift-space 
distortions, and non-linear clustering make theoretical modeling of the 
covariance matrix difficult, and often simulations are used to study
them in detail.
Covariance matrices estimated from a finite number of simulations will
contain noise.
Cosmological parameter estimation requires the inverse of the covariance
matrix to properly weight the measurements.
Matrix inversion is an inherently non-linear operation that is sensitive 
to the noise of all the elements.
\citet{cs06} showed that when the off-diagonal elements of a covariance
matrix are excessively noisy it is better for parameter estimation to
use a diagonal approximation of the covariance.
This reduces the effects of noise, but ignores important information in the 
covariance.

Covariance matrices for large-scale structure measurements are 
often estimated using
the unbiased empirical covariance matrix, $\mathbfss{S}$ (see
equation~\ref{eq:cov}), a close relative of the maximum-likelihood
estimator, $\mathbfss{S}^{(ML)} = \frac {n-1} {n} \mathbfss{S}$.
These estimators work well in the regime where the number of
repeat observations, $n$, is much greater than the number of
parameters measured for each observation, $p$.  However, in
the regimes where $n \sim p$ or $n \ll p$ the covariance matrix
estimates become ill-conditioned and unstable during inversion,
which is necessary for optimal weighting of the data.  
This is an indication that these estimators do not produce good 
approximations of the true underlying covariance matrix in these regimes.
\citet{efron82} provides some insight into the
difference between maximum-likelihood as a {\it summarizer} and
as an {\it estimator}.  Maximum-likelihood is an excellent summarizer
of data in the sense of trying to represent the important statistical
information about a dataset in a small set of numbers.  
Though maximum-likelihood is asymptotically optimal for estimation
in the limit of infinite data, the use of this summary of information 
for the purpose of making estimates with a finite set of data
is not always the best option.  
\citet{stein56} proved that one can construct estimators in high-dimensional 
($d \geq 3$) inference problems that outperform maximum-likelihood 
estimators in the sense of minimizing the {\it total} mean squared error.
Maximum-likelihood produces the best estimates of individual
parameters, but the alternatives can often reduce the error on many
of the parameters while only slightly increasing the error on a few,
resulting in an overall improvement.
\citet{stein56} also showed that the maximum likelihood estimator has 
the best performance among estimators that transform correctly
under translation, implying that any estimator that outperforms
maximum-likelihood will necessarily involve an arbitrary choice.

\citet{ss05} employ a method known as {\it shrinkage estimation}
to construct covariance matrices for functional genomics
measurements in the $n \ll p$ regime.  Their technique optimally combines
a high-dimensional estimate that has little or no bias with a 
low-dimensional estimate that may be biased but has much less variance.
The result minimizes the total mean squared error, which is the sum
of bias (squared) and variance.  
They argue that their method can also perform some amount of regularization,
resulting in a covariance matrix that has a full set of
positive-definite eigenvalues and is well-conditioned (i.e., the ratio
of the largest to smallest eigenvalue is not so large that inversion
becomes unstable).
They employ a lemma from \citet{lw03} to analytically calculate 
the optimal linear combination of the low and high dimensional estimates.

In this paper our goal is to provide a simple recipe for using shrinkage
estimation to improve the covariance matrix of the
matter power spectrum from a limited number of simulations over the
ubiquitous sample covariance estimator.
Our method aims to reduce the total noise
while retaining as much information about real covariance in the
simulations as possible.
Shrinkage estimation achieves this by optimally combining a theoretical
model with the empirical estimate.
We will assess the improvements our method offers by examining the
performance of the covariance matrices through inversion and
use in cosmological parameter estimation.
Although we focus on the matter power spectrum, the shrinkage
technique is relevant for many studies in large-scale structure
and cosmology.

The outline of this paper is as follows:
in Section~\ref{sec:shrink} we introduce shrinkage estimation and
describe its application to covariance matrices.
Section~\ref{sec:toy} applies the shrinkage technique to several toy
problems before moving to a more complicated example involving
galaxy clustering.
We describe our technique for measuring matter power spectra from N-body 
simulations in section~\ref{sec:sims}.
In Section~\ref{sec:results} we construct several estimates of
the power spectrum covariance matrix and compare their performance
by estimating cosmological parameters.
Finally we review our results, make recommendations, and discuss future 
directions of this project in Section~\ref{sec:disc}.

\section{Shrinkage Estimation}
\label{sec:shrink}

\subsection{The Method}
\label{sub:shrink}

Much of this section summarizes the introduction to shrinkage
estimation given in \citet{ss05}.  
Suppose we are estimating a vector $\bmath{\psi}$ (of length $p$)
from a set of $n$ measurements using two different models.  
One of the models has many free parameters and
produces an estimate, $\bmath{u}$, with little (or no) bias, but
the variance may be significant due to the number of free parameters.  
The second model (called the {\it target}) has many fewer (or no) free 
parameters and produces an estimate, $\bmath{t}$, which will have smaller 
variance but may be biased.  
We construct a new estimate, $\bmath{u}^{\star}$, 
from a linear combination of these two models, given by
\begin{equation}
\bmath{u}^{\star} = \lambda \bmath{t} + (1-\lambda) \bmath{u}
\label{eq:ustar}
\end{equation}
where $\lambda \in [0,1]$ is called the {\it shrinkage intensity}.  
The question now becomes how to choose $\lambda$ in an optimal way.  
A common way to optimize an estimator is to minimize the 
expected mean squared error, given by the risk function
\begin{equation}
R(\lambda) = 
\left\langle \sum_{i=1}^{p} (u^{\star}_i - \psi_i)^2 \right\rangle
\label{eq:mse}
\end{equation}
where the angle brackets indicate the expectation value.
\citet{lw03} introduced an analytic solution for the optimal 
shrinkage intensity, $\lambda^{\star}$.
Prior to this solution shrinkage estimation was much less practical
because numerically complicated and expensive methods were necessary
to find the optimal shrinkage intensity.
The analytic solution is
\begin{equation}
\lambda^{\star} = 
\frac {\sum_{i=1}^{p} {\rm Var}(u_i) - {\rm Cov}(t_i,u_i)
- {\rm Bias}(u_i) \langle t_i-u_i \rangle}
{\sum_{i=1}^{p} \langle (t_i-u_i)^2 \rangle}
\label{eq:lstar}
\end{equation}
where ${\rm Var}$, ${\rm Cov}$, and ${\rm Bias}$ are the true
variance, covariance, and bias, respectively.  For a practical
estimator, \citet{ss05} suggest estimating $\hat{\lambda}^{\star}$ as
\begin{equation}
\hat{\lambda}^{\star} = \frac 
{\sum_{i=1}^{p} \widehat{\rm Var}(u_i) - \widehat{\rm Cov}(t_i,u_i)
- \widehat{\rm Bias}(u_i) (t_i-u_i)}
{\sum_{i=1}^{p} (t_i-u_i)^2}
\label{eq:estlstar}
\end{equation}
where $\widehat{\rm Var}$, $\widehat{\rm Cov}$, and $\widehat{\rm Bias}$
are the unbiased sample
estimates of ${\rm Var}$, ${\rm Cov}$, and ${\rm Bias}$,
respectively.  
The ${\rm Bias}$ term can be ignored if $\bmath{u}$ is an 
unbiased estimator.

\tabletoy

\subsection{Application to Covariance Matrices}
\label{sub:shrinkcov}

We now specialize the shrinkage estimation technique to covariance matrices.  
Suppose we have $n$ sets of data and we measure a data vector 
$\bmath{x}$ of length $p$ for each of them.
Let $x^{(k)}_i$ be the $k^{\rm th}$ (of $n$ total) observation
of the $i^{\rm th}$ (of $p$ total) element of the data vector.  
The estimated empirical mean is given by 
$\overline{x}_i = \frac {1} {n} \sum_{k=1}^{n} x^{(k)}_i$.
We define
\begin{eqnarray}
\label{eq:wkij}
W^{(k)}_{ij} & = & 
(x^{(k)}_i - \overline{x}_i)(x^{(k)}_j - \overline{x}_j), \\
\label{eq:wij}
\overline{W}_{ij} & = & 
\frac {1} {n} \sum_{k=1}^{n} W^{(k)}_{ij}
\end{eqnarray}
and write the unbiased empirically estimated covariance matrix
of the data, $\mathbfss{S}$, as
\begin{equation}
S_{ij} = \widehat{\rm{Cov}}(x_i,x_j) 
= \frac {n} {n-1} \overline{W}_{ij}.
\label{eq:wcov}
\end{equation}
Explicit substitution results in the usual
\begin{equation}
S_{ij} = \frac {1} {n-1} \sum_{k=1}^{n} 
(x^{(k)}_i - \overline{x}_i)(x^{(k)}_j - \overline{x}_j).
\label{eq:cov}
\end{equation}
Similarly we can estimate the covariance of the elements
of the covariance matrix of the data, given as
\begin{equation}
\widehat{\rm{Cov}}(S_{ij},S_{lm}) 
= \frac {n} {(n-1)^3} \sum_{k=1}^{n}
(W^{(k)}_{ij} - \overline{W}_{ij}) 
(W^{(k)}_{lm} - \overline{W}_{lm})
\label{eq:covcov}
\end{equation}
with the variance of an individual entry given by
$\widehat{\rm{Var}}(S_{ij}) = \widehat{\rm{Cov}}(S_{ij},S_{ij})$.
For shrinkage estimation we let $\mathbfss{S}$ take the role of 
$\bmath{u}$ and supply a target covariance matrix $\mathbfss{T}$ to take 
the role of $\bmath{t}$.
The optimal shrinkage intensity and resulting covariance 
matrix $\mathbfss{C}$ are given by
\begin{eqnarray}
\label{eq:lshat}
\hat{\lambda}^{\star} & = & \frac
{\sum_{i,j} \widehat{\rm Var}(S_{ij})
- \widehat{\rm Cov}(T_{ij},S_{ij})}
{\sum_{i,j} (T_{ij} - S_{ij})^2}, \\
\label{eq:scov}
\mathbfss{C} & = & \hat{\lambda}^{\star}\mathbfss{T} + 
(1-\hat{\lambda}^{\star}) \mathbfss{S}.
\end{eqnarray}
If the $\hat{\lambda}^{\star}$ estimate is greater than
one, then $\hat{\lambda}^{\star} = 1$ is enforced, implying that only
the target matrix is used.  If the $\hat{\lambda}^{\star}$ estimate
is less than zero, then $\hat{\lambda}^{\star} = 0$ is enforced,
implying that only the empirical covariance matrix is used.
The numerator of equation~\ref{eq:lshat} implies that as the variances 
of the elements of $\mathbfss{S}$ decrease 
(e.g., approaching the $n \gg p$ regime)
the shrinkage estimate smoothly approaches the empirical covariance.
The denominator in equation~\ref{eq:lshat} also ensures that if the
chosen target matrix is very different from the empirical covariance
then the estimator will tend to the empirical covariance.  
Thus an inappropriate choice of target should not make the final results 
any worse, but there will be little gain in efficiency.  

The $\widehat{\rm Cov}(T_{ij},S_{ij})$ term in the numerator of
equation~\ref{eq:lshat} accounts for the fact that $\mathbfss{S}$ and
$\mathbfss{T}$ are estimated from the same data.  If $\mathbfss{T}$
is fixed then the term is zero.  If some of the elements of $\mathbfss{T}$
are taken directly from $\mathbfss{S}$ then the $\widehat{\rm Cov}$
term exactly cancels the $\widehat{\rm Var}$ term and those elements
do not affect the estimate of $\hat{\lambda}^{\star}$.
Tab. 2 of \citet{ss05} shows a variety of worked examples for
common targets that may or may not depend on the empirically estimated
covariance.  In their examples they ignore moments of higher
order than $\widehat{\rm Var}(S_{ij})$.

\section{Toy Examples}
\label{sec:toy}

\figtoy

Before tackling large-scale structure measurements
we present the application of shrinkage estimation to toy models.
In Section~\ref{sub:toymean} we employ shrinkage to estimate
the mean of a noisy vector.
For that example we can calculate the expected shrinkage intensity
analytically.
In Section~\ref{sub:toycov} we present a toy example of covariance
estimation.

\subsection{Estimating the Mean of a Noisy Vector}
\label{sub:toymean}

In our first toy example of the shrinkage technique we will estimate
the mean of a vector from a set of noisy realizations using the
formalism from Section~\ref{sub:shrink}.
Given an input mean vector, $\bmath{\psi}$, the distribution for
an element of the noisy realization is $X_{i} = N(\psi_i,\sigma_i^2)$,
a normal distribution with mean $\psi_i$ and variance $\sigma_i^2$.
The $i^{th}$ element (of $p$ total) of the $k^{th}$ realization
(of $n$ total) is written $x_i^{(k)}$, and our high-dimensional
estimate of the mean is given by
\begin{equation}
\bmath{u} = \frac {1} {n} \sum_{k=1}^{n} \bmath{x}^{(k)}.
\label{eq:u}
\end{equation}
Given a fixed target, $\bmath{t}$, we will calculate the expected value 
for the optimal shrinkage intensity, $\langle \lambda^{\star} \rangle$,
as given by equation~\ref{eq:lstar}.  If we assume the numerator and
denominator are uncorrelated we can write
\begin{equation}
\langle \lambda^{\star} \rangle = 
\left \langle 
\frac {\sum_{i=1}^{p} {\rm Var}(u_i)}
{\sum_{i=1}^{p} \langle (t_i-u_i)^2 \rangle}
\right \rangle
\approx
\frac {\sum_{i=1}^{p} \left \langle {\rm Var}(u_i) \right \rangle}
{\sum_{i=1}^{p} \left \langle (t_i-u_i)^2 \rangle \right \rangle}.
\label{eq:explstar}
\end{equation}
From our definitions we know that
\begin{eqnarray}
\label{eq:expx}
\langle x_i^{(k)} \rangle & = & \psi_i, \\
\label{eq:expx2}
\langle x_i^{(k)} x_j^{(l)} \rangle & = &
\psi_i \psi_j + \delta_{ij} \delta_{kl} \sigma_i^2
\end{eqnarray}
which we can use to find the expectation values of our estimators
\begin{eqnarray}
\label{eq:expu}
\langle u_i \rangle & = &
\left \langle \frac {1} {n} \sum_{k=1}^{n} x_i^{(k)} \right \rangle
= \psi_i,\\
\label{eq:expu2}
\langle u_i^2 \rangle & = & \left \langle 
\frac {1} {n} \sum_{k=1}^{n} x_i^{(k)}
\frac {1} {n} \sum_{l=1}^{n} x_i^{(l)}
\right \rangle
= \psi_i^2 + \frac {1} {n} \sigma_i^2.
\end{eqnarray}
Fixing the values for $\psi_i$, $\sigma_i$, and $t_i$ to be
$\psi$, $\sigma$, and $t$, respectively, we predict the shrinkage
intensity to be
\begin{equation}
\langle \lambda^{\star} \rangle \approx \frac {\sigma^2}
{n(t-\psi)^2 + \sigma^2} .
\label{eq:explstarfinal}
\end{equation}
This formula displays the behavior we expect.  As $n$ becomes large,
$\langle \lambda^{\star} \rangle$ 
approaches zero and we will rely mostly on the empirical
estimate of the mean.  As $t$ approaches $\psi$, 
$\langle \lambda^{\star} \rangle$ 
approaches unity as we have chosen the perfect target.
We can also calculate the expected mean squared errors for the
high-dimensional, low-dimensional, and shrinkage estimators 
and find they are given by
\begin{eqnarray}
\label{eq:mseu}
\langle {\rm MSE}(\bmath{u}) \rangle & = & p \frac {\sigma^2} {n},\\
\label{eq:mset}
\langle {\rm MSE}(\bmath{t}) \rangle & = & p(\psi-t)^2,\\
\label{eq:mseustar}
\langle {\rm MSE}(\bmath{u}^{\star}) \rangle & = &
p \left [ \langle \lambda^{\star} \rangle^2 (\psi-t)^2 +
(1-\langle \lambda^{\star} \rangle^2) \frac {\sigma^2} {n} \right ],
\end{eqnarray}
respectively.

We generated simulations to check the
validity of our analytic predictions.  Each simulation starts by
generating the $n$ realizations $\bmath{x}^{(k)}$.  The sample mean
$\bmath{u}$ is calculated for the set of realizations and used with the
target $\bmath{t}$ to generate the shrinkage estimate $\bmath{u^{\star}}$.
The shrinkage intensity, $\hat{\lambda}^{\star}$,
for the set of realizations is calculated using
equation~\ref{eq:estlstar} and jackknife resampling of equation~\ref{eq:u}
is used to estimate the $\widehat{\rm Var}(u_i)$ term.
We generated 100 simulations for each set of input parameters we
tested and the results are shown in Table~\ref{tab:toy}.
The results of our simulations match the analytical predictions
reasonably well and also show that the shrinkage estimator outperforms
both the empirical estimate and the target in a mean squared error sense.

\subsection{Estimating the Covariance}
\label{sub:toycov}

Our second toy model applies the formalism of Section~\ref{sub:shrinkcov}
for shrinkage estimation of a covariance matrix.
In this case the analytical predictions would be prohibitively
tedious, so we present only the results of simulations.
Each simulation starts by generating the realizations,
$\bmath{x}^{(k)}$, where the distribution of each element is
$X_i^{(k)} = N(0,\sigma^2)$.
From each set of realizations we construct two estimates of the
covariance matrix.
The Monte Carlo (MC) method uses the standard empirical covariance estimate 
as given by equation~\ref{eq:cov}.
We also compute a shrinkage version of the Monte Carlo estimator (MC+S) from
equation~\ref{eq:scov} using the identity matrix as a target.
The two estimators are compared by inspecting the eigenvalue spectra
and by computing the mean squared error between the
estimate and the known true covariance matrix,
which is $\sigma^2\mathbfss{I}$ where $\mathbfss{I}$ is the identity matrix.
The mean squared error between two matrices $\mathbfss{A}$ and $\mathbfss{B}$
is given by the Frobenius norm of the difference,
\begin{equation}
||\mathbfss{A}-\mathbfss{B}||^2_{F}
= \sum_{i,j} |A_{ij}-B_{ij}|^2.
\label{eq:frobenius}
\end{equation}

For our simulations we fixed $p=18$ and $\sigma=1.1$ and compared
the performance of the covariance estimators as a function of the
number of realizations, $n$.  We ran 100 simulations for each value
of $n$ and averaged the results.
For $n = 40, 400, 4000$ the shrinkage intensities were
$\hat{\lambda}^{\star} = 0.92 \pm 0.08$, $0.61 \pm 0.06$, $0.136 \pm 0.006$.
Figure~\ref{fig:toy} shows the results for the two estimators as well
as the results when only using the target.
The shrinkage estimator gives a mean squared error that is equal to or
less than both the empirical covariance and the target by itself for all $n$.
The eigenvalue spectrum for the shrinkage estimator is also closer to the
correct spectrum than for the empirical covariance at fixed $n$.
This indicates that using the shrinkage estimator is in some sense
equivalent to using more simulations.

\section{Simulations and Measurements}
\label{sec:sims}

We divide the Hubble Volume \citep{hv} $\Lambda$CDM simulation into 
4096 sub-volumes, each with a sidelength of $187.5~{\rm Mpc/h}$.  
We measure the power spectrum in each sub-volume using a simple code 
that implements the same basic algorithm as presented in \citet{espice}
and employs a Fast Fourier Transform (FFT).
The FFT was performed on a $64^3$ grid and the power
spectrum was measured in 18 logarithmically-spaced bins from 
$k=0.0367$ to $k=0.920$.  
For plotting and model comparison purposes we calculate the average $k$ 
from the actual modes in each bin for the bin centers.  
The power spectrum calculated on a pixelized grid is the product of
the true power spectrum with (the square of) the Fourier transform of
the pixel window function.
Our measured power spectrum can be written as
$P_{\rm meas}(\bmath{k}) = P(\bmath{k})/|\tilde{W}_p(\bmath{k})|^2$
where the Fourier transform of the pixel window function, $\tilde{W}_p$,
is given by
\begin{eqnarray}
\label{eq:f}
\tilde{f}(k) & = & 
\frac {{\rm sin}(\pi k / (2 \pi/L))} {\pi k / (2 \pi/L)},\\
\label{eq:w}
|\tilde{W}_p(\bmath{k})|^2 & = & \tilde{f}^2(k_x)
\tilde{f}^2(k_y) \tilde{f}^2(k_z),
\end{eqnarray}
and $L$ is the length of the side of a pixel.  
Our pixels are $2.93~{\rm Mpc/h}$ on each side.

We used {\sevensize CAMB} \citep{camb} to calculate a model transfer function 
and matter power spectrum using cosmological parameter values to match 
the Hubble Volume \citep{hv} $\Lambda$CDM simulation: 
$\Omega_m = 0.3$, 
$\Omega_{\Lambda} = 0.7$,
$h = {\rm H}_0/100~{\rm km~s^{-1} Mpc^{-1}} = 0.7$,
$\Omega_bh^2 = 0.0196$,
and $n_s = 1$.
We normalized the resulting power spectrum, $P_{\rm CAMB}(k)$, to have a 
(linear) $\sigma_8 = 0.9$ and then applied the non-linear 
{\sevensize HALOFIT} \citep{halofit} correction.
Thus the normalization matches linear theory for large scales
(low $k$), but the shape includes non-linear clustering corrections
at smaller scales (higher $k$).

The measured power spectrum is the convolution of the true power
spectrum with the squared Fourier transform of the survey window function.
When testing a model for the power spectrum we must perform this
convolution before comparing it to our measurement.  The
convolved power spectrum, $P^W$, is given by
\begin{equation}
P^W(\bmath{k}) = \int d\bmath{q}~P(\bmath{k}-\bmath{q})
|\tilde{W}_s(\bmath{q})|^2
\end{equation}
where $P$ is the input (theoretical) power spectrum.
Because the survey is the same shape as our pixels we can write
the (normalized) survey window function in terms of the pixel window 
function as $|\tilde{W}_s|^2 = (L/2\pi)^6|\tilde{W}_p|^2$ 
where $L$ is now the survey sidelength.
To simplify the convolution we used the
spherically averaged transform of the window function (calculated
via Monte Carlo) and reduced the convolution of the spherically
symmetric power spectrum to a two-dimensional integral, given by
\begin{equation}
P^W(k) \approx 2\pi \int_0^{\infty}dq~q^2 |\tilde{W}_s(q)|^2
\int_{-1}^{1}dx~P(\sqrt{k^2+q^2-2kqx}).
\end{equation}
In practice the integral over $dx$ was done with Romberg integration
and the integration over $dq$ was performed with the extended
trapezoidal rule at the $k$ values where we had calculated the
spherically averaged $|\tilde{W}_s(\bmath{k})|^2$.  
Power spectrum values inside the integral were evaluated using cubic 
spline interpolation in $({\rm log}~k, {\rm log}~P)$ and power law 
extrapolation at low and high $k$.  
Fig.~\ref{fig:pk} shows the input theoretical power spectrum 
$\left( P_{\rm CAMB} \right)$, 
the convolved model power spectrum $\left( P^W_{\rm CAMB} \right)$, 
and the average of the measured power spectra from all of the sub-volumes
$\left( \langle P_{\rm meas} \rangle \right)$.
The inset shows the spherically averaged survey window
function.

\figpk

We wish to evaluate our covariance estimates in the context of
cosmological parameter fitting.  Given the relatively small volume of 
each of our sub-volumes we decided to fit for only the power spectrum
normalization in each sub-volume, parameterized by $\sigma_8$ (linear).
We write the log-likelihood (for a fixed covariance matrix) as
\begin{equation}
\bmath{d} (\sigma_8) = P_{\rm meas}(k) - \left ( \frac {\sigma_8} 
{0.9} \right )^2 P^W_{\rm CAMB}(k,\sigma_8=0.9 ),
\label{eq:lnld}
\end{equation}
\begin{equation}
{\rm log} {\mathcal L} (\sigma_8) \propto - \frac {1} {2} 
\bmath{d}^T \mathbfss{C}^{-1} \bmath{d}
= - \frac{1}{2} \chi^2
\label{eq:lnl}
\end{equation}
where $\mathbfss{C}$ is the covariance matrix being tested.  
We use $\sigma_8$ to fit for the amplitude of the power
spectrum in the linear regime, and we assume that the difference in
shape in the non-linear regime is minimal for values of $\sigma_8$
that are close to our fiducial model of $\sigma_8=0.9$.
We numerically find the maximum likelihood value, $\hat{\sigma}_8$,
and define the upper and lower one sigma error bars, $\pm \Delta$, 
where
\[
{\rm log} {\mathcal L}(\hat{\sigma}_8 \pm \Delta) 
- 
{\rm log} {\mathcal L}(\hat{\sigma}_8) 
= -1/2.
\]

Our likelihood analysis assumes that the bandpower measurements of
the power spectrum are normally distributed.  For most of our bandpowers
this is a good approximation, but there are several for which an
offset lognormal distribution \citep{bjk00} would be more accurate.
This could cause some bias in our recovered value of $\hat{\sigma}_8$,
but it should not affect our assessment of the relative performance
of different covariance estimators.  The covariance matrices for
the normal and offset lognormal cases are related by a change
of variables.

\section{Covariance Matrix Estimates}
\label{sec:results}

In this section we compare the performance of several techniques
for estimating the covariance matrix of our power spectrum measurements.
As a baseline comparison we calculate the covariance matrix from all
4096 sub-volumes using equation~\ref{eq:cov}.  For each method of
covariance matrix estimate we measure $\hat{\sigma}_8$ and
$\Delta$ using the power spectra measured from the sub-volumes
and show the distributions of the $\hat{\sigma}_8$ and $\Delta$ values.
The mean and standard deviations of those quantities are presented
in Table~\ref{tab:s8}.

\tableresults

\subsection{Reference}
\label{sub:ref}

As a reference we estimate the covariance matrix of our power spectrum
measurement by applying equation~\ref{eq:cov} to our measurements from
all 4096 sub-volumes.  
There are $18 \times 19/2 = 171$ independent elements in the covariance 
matrix, thus we are in the regime where we
have many more realizations than elements to be estimated and the
usual covariance estimator should give reasonable results.
The solid black line in
Fig.~\ref{fig:s8mc} is a histogram of the results of estimating
$\hat{\sigma}_8$ using the reference covariance matrix and the
power spectra measured from each of the 4096 sub-volumes.  
The upper panel shows the distribution of $\hat{\sigma}_8$ and the
lower panel shows the distribution of the error bar estimates
(absolute value of both upper and lower).
The mean and standard deviation of these histograms is presented
in Table~\ref{tab:s8}.
The agreement between the width of the best-fit
distribution, $\sigma_{\hat{\sigma}_8}$,
and the mean error bar estimate, $\langle \Delta \rangle$, 
indicates that the covariance
matrix is properly estimating the likelihood distribution.
The width of the error bar distribution, $\sigma_\Delta$,
is small, indicating that the error bar estimate is usually
very close to the correct value.
The mean of the maximum-likelihood estimates, 
$\langle \hat{\sigma}_8 \rangle$, 
is $0.870$ instead of our known input value of $0.9$, 
but we know that our modeling of the power
spectrum into the non-linear regime is not perfect so this small
offset is not worrisome for our purposes.
For the remainder of this section we assume that the results using the 
reference covariance matrix are a good approximation to those
that would be obtained using the true underlying covariance matrix.
See Section~\ref{sec:disc} for further discussions.

\subsection{Monte Carlo}
\label{sub:mc}

\figsmc

Next we test covariance matrices estimated with 
equation~\ref{eq:cov} but using a small number of sub-volumes,
which we call the Monte Carlo method.
We use sets of 40 (randomly chosen, non-overlapping) 
sub-volumes to test the regime where we have more simulations
than diagonal elements of the covariance matrix (18), but fewer
simulations than independent elements (171).
From 4096 sub-volumes we can create 102 separate covariance matrix
estimates.
To obtain smooth histograms in Fig.~\ref{fig:s8mc} we test each covariance
matrix estimate against 40 randomly chosen $P(k)$ measurements
from other sub-volumes.
The statistics in Table~\ref{tab:s8} are calculated
using one randomly chosen $P(k)$ measurement per covariance matrix.
The statistics do not depend on how many randomly chosen $P(k)$
measurements are used for each covariance matrix estimate.
The upper panel of Fig.~\ref{fig:s8mc} shows that the distribution
of $\hat{\sigma}_8$ is too wide, indicating that a parameter
analysis using a covariance matrix estimated with this method will
often return a value far from the mean.
The lower panel of Fig.~\ref{fig:s8mc} shows that the error bar
is typically underestimated by $\sim 25\%$.
The width of the estimated error
bar distribution is also much wider than for the reference covariance
matrix estimate.  These effects are the result of using a very noisy
estimate of the covariance matrix.

\subsection{Monte Carlo + Shrinkage}
\label{sub:mcs}

\figcov

Our first test of the shrinkage approach is to apply shrinkage
estimation to the Monte Carlo method described in the previous
section.  First we need to choose a target covariance matrix,
$\mathbfss{T}$.  
In linear theory we expect the covariance matrix of the power spectrum
to be diagonal.  Off-diagonal terms arise in practice from
the survey window function and non-linear clustering effects.  
We use a diagonal target matrix to simulate a situation where we have some 
idea about the structure of the covariance matrix but we know our model is 
not exact.  Our target matrix takes the form
\begin{equation}
T_{ij} = \left\{
\begin{array}{ll}
0 & i \neq j \\
2 [P^W_{\rm CAMB} (k_i)]^2/N_i & \langle k_i \rangle \leq 0.14~{\rm h/Mpc}\\
S_{ii} & \langle k_i \rangle > 0.14~{\rm h/Mpc}
\end{array}
\right.
\label{eq:target}
\end{equation}
where we use a different method in the linear and non-linear
regimes.  For bins in the linear regime we use our convolved model 
for the power spectrum, $P^W_{\rm CAMB}(k_i)$, and the number 
of $k$ modes in each bin, $N_i$, to predict the covariance
\citep*[e.g.,][]{hrs}.  In the non-linear regime we use the diagonal
of the empirically estimated covariance from the 40 sub-volumes.
Fig.~\ref{fig:cov} shows the diagonal elements of the reference
covariance matrix, the linear theory model, and the 102 target
matrices.  Inset is the reference correlation matrix,
$R_{ij} = C_{ij}/\sqrt{C_{ii}~C_{jj}}$,
showing that the covariance matrix is strongly diagonal until
well into the non-linear regime.  
Our results are robust to changes in the non-linear cutoff
by several bins in either direction.

We calculate the optimal shrinkage intensity, $\hat{\lambda}^{\star}$, 
for each of the 102 Monte Carlo estimates, $\mathbfss{S}$,
according to equation~\ref{eq:lshat}.  We apply the
$\widehat{\rm Cov}(T_{ij},S_{ij})$ term to the diagonal elements of 
$\mathbfss{T}$ that are taken from $\mathbfss{S}$.
We find values for $\hat{\lambda}^{\star}$ distributed evenly
between $0.1$ and $1.0$ (see Fig.~\ref{fig:lambda}).
We produce each of our 102 new estimates of the covariance matrices, 
$\mathbfss{C}$, from a linear combination of $\mathbfss{S}$ and 
$\mathbfss{T}$ according to equation~\ref{eq:scov}.  
We perform the same tests as
described in section~\ref{sub:mc} and compare the results
in Fig.~\ref{fig:s8mc} and Table~\ref{tab:s8}.
The most striking result is that the maximum-likelihood estimates,
$\hat{\sigma}_8$, follow a very similar distribution to that for
the reference matrix, indicating that the parameter values are now
correctly estimated.
The error bars are still underestimated, but the distribution is
very similar to that for the normal Monte Carlo estimator. 

Fig.~\ref{fig:s8mc} and Table~\ref{tab:s8} also show results
using only the diagonal target.  The values of $\hat{\sigma}_8$
follow the correct distribution, indicating that the estimated
parameter values are fine, but the error bars are much more
severely underestimated.  This is expected because our target
matrix is diagonal and we are using information from far enough into
the non-linear regime to know that we are missing some important
covariance.
It is now clear that the estimated error bar distribution of the shrinkage
estimator is a combination of the Monte Carlo and target distributions.
The shrinkage intensity can serve as a proxy for whether the estimated
error bars are likely to be similar to those for the Monte Carlo
or the target.  See Section~\ref{sec:disc} for further discussions.

In summary, the shrinkage of the empirically estimated covariance
against our target matrix outperforms either matrix by itself.
Using just the empirically estimated covariance brings in too much
noise which causes error in the estimation of $\hat{\sigma}_8$
itself.  Using only the diagonal target mitigates the noise problems,
but ignores important covariance.  The shrinkage estimator
uses the best aspects of both, keeping the part of the covariance
that is well estimated but drastically reducing the total amount
of noise.

\subsection{Jackknife}
\label{sub:jk}

\figsjk

Recently a resampling technique know as the jackknife method has been
used to estimate covariance matrices for large-scale structure
measurements from the data set itself. 
The method works by dividing the data volume
into $n$ cells of roughly the same size and recalculating the
measurement $n$ times, each time with a different cell left out.
The variance between the measurements can be adjusted to try and
calculate the variance corresponding to the entire volume.  In practice
one replaces equation~\ref{eq:wkij} with
\begin{equation}
W^{(k)}_{ij} = \frac {(n-1)^2} {n}
(x^{(k)}_i - \overline{x}_i)(x^{(k)}_j - \overline{x}_j)
\label{eq:jkwkij}
\end{equation}
and then calculates the covariance matrix with equation~\ref{eq:wcov},
resulting in the usual
\begin{equation}
S_{ij} = \frac {n-1} {n} \sum_{k=1}^{n} 
(x^{(k)}_i - \overline{x}_i)(x^{(k)}_j - \overline{x}_j).
\label{eq:covjk}
\end{equation}
We divided each sub-volume into $3^3 = 27$ cells and modified our
code to calculate the power spectrum with one cell removed.  Our
code incorporates a volume correction, the lowest order edge
correction in Fourier space.  For each of the 4096 
sub-volumes we estimate $\hat{\sigma}_8$ and $\Delta$ using the power
spectrum and the jackknife covariance matrix from the same
sub-volume.  The results are compared to the reference case in
Fig.~\ref{fig:s8jk} and listed in Table~\ref{tab:s8}.

The distribution of $\hat{\sigma}_8$ 
is much wider than for the reference covariance matrix, indicating that
the noise in the covariance estimate causes incorrect parameter
estimation.  This is similar to the result for the Monte Carlo method.
The jackknife estimates of $\hat{\sigma}_8$ also peak at a noticeably lower 
value than for the reference covariance, though the two
histograms are in roughly $1 \sigma$ agreement given the width of
the distribution for the jackknife case.  
The error bars estimated
in the jackknife case are typically underestimated by a factor
of almost three compared to the reference covariance matrix and nearly
an order of magnitude compared to the actual width of the
jackknife distribution of $\hat{\sigma}_8$.

\subsection{Jackknife + Shrinkage}
\label{sub:jks}

Our final method of estimating the covariance matrix
applies shrinkage to the jackknife estimator to see if we can achieve
enhanced robustness.  
We use the same method to construct a target matrix as described in
section~\ref{sub:mcs}, using the diagonal of the jackknife
estimated covariance matrix in the non-linear regime.
We calculate the shrinkage intensity, $\hat{\lambda}^{\star}$,
and covariance estimate
for each of the 4096 covariance matrices as described in
section~\ref{sub:shrinkcov}, but substituting
equation~\ref{eq:jkwkij} for equation~\ref{eq:wkij} throughout.  
We find values for $\hat{\lambda}^{\star}$ distributed evenly
between $0.0$ and $1.0$ (see Fig.~\ref{fig:lambda}).
We run the same tests as described in section~\ref{sub:jk} and
the results are shown in Fig.~\ref{fig:s8jk} and Table~\ref{tab:s8}.  

As with the shrinkage version of the Monte Carlo estimator, the
shrinkage version of the jackknife estimator shows significant
improvement in the actual estimated parameter, $\hat{\sigma}_8$.
However, the central value and width are not quite as good as for
the reference case.  There is some improvement
in the estimation of the error bar, though the error bars are still
systematically underestimated by a factor of roughly two
compared to the reference.

Fig.~\ref{fig:s8jk} and Table~\ref{tab:s8} also show the results
of estimating $\hat{\sigma}_8$ and $\Delta$ using only the diagonal
targets used in the shrinkage version of the jackknife estimator.
Again, the diagonal target matrix does well for estimating
$\hat{\sigma}_8$ due to the lack of noise, but it gives the worst
estimates of the error bars.

In this case, the shrinkage version of the jackknife estimator
did the best job of estimating the error bars, and it was only
slightly worse than the diagonal approximation at recovering
the distribution of $\hat{\sigma}_8$.  Again, shrinkage estimation
is doing an excellent job of keeping information about covariance
while reducing the total noise.

\section{Discussions and Conclusions}
\label{sec:disc}

We have introduced shrinkage as a technique for improving estimates
of the covariance matrix for power spectrum measurements.
We tested our methods on dark matter simulations and showed
improvement over the empirically estimated covariance matrix from
a limited number of simulations or jackknife resamplings.
In order to clearly assess the potential improvement from using
shrinkage estimation, we chose an intentionally difficult scenario
where traditional methods of estimating the covariance were unlikely
to yield satisfactory results.  
All of these methods would perform better if we allowed ourselves
more simulations per Monte Carlo estimate or if we did not push as
far into the non-linear clustering regime.  The shrinkage technique
would still outperform the other methods, but perhaps the differences
would be less obvious.

A good estimate of the covariance matrix of a power spectrum
measurement is essential for extracting cosmological information via
parameter fitting.  Including the covariance between different bins
is a good step towards properly estimating the confidence intervals 
on cosmological parameters.  However, the increased number of free
parameters of a full covariance estimate (as opposed to a diagonal
approximation) can cause the covariance estimate to be noisy if only
a relatively small number of simulations are available.  This noise
can adversely affect the estimate of the parameter itself.  A diagonal 
approximation to the covariance can be more easily constrained with a 
limited number of simulations, leading to better estimates of the 
parameter values.  However, the confidence intervals can be severely
underestimated if actual covariance is ignored.  Neither alternative
is appealing.  If a similar measurement was performed with the
two-point correlation function, the Fourier dual of the power spectrum,
a full covariance matrix is especially important as bins will be strongly
correlated, even in the linear clustering regime.
Realistic survey geometries will also cause additional covariance on large
scales for the power spectrum.

Shrinkage estimation is an optimal way of combining a model with many 
degrees of freedom and a model with few degrees of freedom
to minimize the total error on the covariance estimate.  
In our example the shrinkage versions of the Monte Carlo and jackknife
estimators clearly outperformed their counterparts without shrinkage,
with the shrinkage version of the Monte Carlo estimator producing
the best results.
The lemma of \citet{lw03} as employed by \citet{ss05} allows a
mathematically and numerically simple way of calculating the optimal
shrinkage intensity.
This means that there is minimal addition work required to use a
shrinkage version of a covariance estimator.
Shrinkage estimation can result in a massive improvement in the limit of 
a small number of simulations and will not adversely affect the 
covariance estimate in the limit of a large number of simulations.
For these reasons we always recommend the use of the shrinkage versions 
of covariance estimators in all regimes.

We briefly investigated the effects of shrinkage estimation in the
limit of a large (though not infinite) number of simulations.
We applied shrinkage estimation to our reference covariance matrix
estimated from all 4096 sub-volumes using the target from
equation~\ref{eq:target} and found an optimal shrinkage intensity
$\hat{\lambda}^{\star} = 0.0096$.  This number is the same order as
the relative noise we expect in each element of the matrix,
$1/\sqrt{4096} = 0.0156$.  We then calculated the eigensystems of
both matrices.  The dot products of the corresponding eigenvectors
always exceeded $0.996$, indicating that
they are essentially identical.  The (sorted) eigenvalue spectra
are shown in Fig.~\ref{fig:eigen}.  The eigenvalues are the same
to within $1\%$ for the first 10 eigenmodes.  After the tenth eigenmode
the eigenvalues from the reference matrix become increasingly smaller
compared to the shrinkage version.  By the final eigenmode the difference
is $\sim 50\%$.  The shrinkage version of the reference matrix
should be a more accurate estimate of the true underlying covariance
matrix.  
The non-linear nature of matrix inversion can cause errors $\gg 1\%$ even 
when individual elements of the covariance matrix are estimated to $\sim 1\%$.
We ran our parameter estimation test using the shrinkage version of the 
reference matrix and found that $\langle \hat{\sigma}_8 \rangle$ moved by 
less than $0.5\%$.
This is small compared to the width of the distribution, which is $\sim 5\%$.  
The average minimum $\chi^2$ did improve from $52.4$ to $41.1$ with the 
shrinkage version of the covariance matrix, though this is still large 
for $18 - 1 = 17$ degrees of freedom.
The remaining discrepancy is dominated by bias from problems with modeling
the power spectrum into the non-linear regime or power loss in the
simulation at smaller scales due to low resolution, not a grossly
inaccurate estimate of the variances.
The amplitude is mainly sensitive to smooth eigenmodes, which have
large eigenvalues, so there is little change in the estimated value.
Parameters that are more sensitive to the shape of the power spectrum may 
be more sensitive to the lower eigenvalue modes and show more than a
$1\%$ change.
The impact of these differences could be estimated with a study of
the information content of the power spectrum covariance 
in terms of cosmological parameter confidences (i.e., \citealt{ns07}), 
but this is beyond the scope of this paper.

\figeigen

We employed a very simple diagonal target matrix in this paper, 
but better targets can clearly improve the efficiency of the shrinkage
technique.  A much more realistic model for covariance on small scales
could be constructed using the halo model.  For realistic measurements
it may also be advantageous to model some of the effects of survey
geometries, redshift-space distortions, and clustering bias.
Targets that depend on a small number of free parameters may be very
useful for some of these effects (e.g., clustering bias).
Targets can also be developed for a wide range of large-scale
structure measurements in addition to the power spectrum.
The exploration of more sophisticated targets is beyond the scope of
this paper and is left to future studies.

Ultimately we would like to develop more diagnostics of the
performance of our covariance estimates.  Fig.~\ref{fig:lambda}
shows the estimated error bar, $\Delta$, as a function of the
shrinkage intensity, $\hat{\lambda}^{\star}$, for the shrinkage
versions of the Monte Carlo and jackknife estimators.  There is
clearly some correlation for the shrinkage version of the
Monte Carlo estimator, so knowledge of $\hat{\lambda}^{\star}$
could help one gauge how much the error bars are underestimated.
The exploration of such diagnostics should proceed as better
targets are developed.

\figlambda

The difficulties in estimating the power spectrum covariance
matrix in the context of making precision cosmological
measurements are of even greater concern for higher-order
clustering measurements.  Higher-order clustering measurements
have a configuration space with more degrees of freedom than the power 
spectrum (or two-point correlation function).  Even a lower resolution
measurement will have more bins and a much larger
covariance matrix, and noise will cause larger deviations in the
inverse matrix.  Theoretical modeling of the covariance matrix
for an N-point correlation function generally involves correlations
up to the 2N-point \citep[e.g.,][]{higher}, making the models more uncertain.
The ability to optimally combine simulations and a theoretical
model with a small number of free parameters will make dramatic
improvements.  Shrinkage estimators could also be used for covariance 
matrices of measurements outside of large-scale structure, including the 
cosmic microwave background power spectrum.  
Finally, we note that Section~\ref{sub:shrink} makes no specific references 
to covariance matrices and that shrinkage is a general estimation technique.
We are studying additional applications of shrinkage estimation for
cosmological measurements.

\section*{Acknowledgments}
The authors thank Mark Neyrinck and Gang Chen for discussions about
the covariance matrix of the power spectrum and the effects of noise.
The authors are grateful for support NASA grant NNG06GE71G
and NSF grant AMS04-0434413.

\end{document}